# Active spintronic-metasurface terahertz emitters with tunable chirality

Changqin Liu[1,2†], Sheng Zhang[1†], Shunjia Wang[1†], Qingnan Cai[1†], Peng Wang[1], Chuanshan Tian[1], Lei Zhou[1*], Yizheng Wu[1*], Zhensheng Tao[1*]

[1] *Department of Physics and State Key Laboratory of Surface Physics, Fudan University, Shanghai 200433, China*

[2]*Shanghai Research Center for Quantum Sciences, Shanghai 201315, China*

[†]These authors contributed equally to this work.

*Corresponding authors: Dr. Lei Zhou, phzhou@fudan.edu.cn; Dr. Yizheng Wu, wuyizheng@fudan.edu.cn; Dr. Zhensheng Tao, ZhenshengTao@fudan.edu.cn.

## Abstract

The ability to manipulate the electric-field vector of broadband terahertz waves is essential for applications of terahertz technologies in many areas, and can open up new possibilities for nonlinear terahertz spectroscopy and coherent control. Here, we propose a novel laser-driven terahertz emitter, consisting of metasurface-patterned magnetic multilayer heterostructures. Such hybrid terahertz emitters can combine the advantages of spintronic emitters for being ultrabroadband, efficient and flexible, as well as those of metasurfaces for the unique capability to manipulate terahertz waves with high precision and degree of freedom. Taking a stripe-patterned metasurface as an example, we demonstrate the generation of broadband terahertz waves with tunable chirality. Based on experimental and theoretical studies, the interplay between the laser-induced spintronic-origin currents and the metasurface-induced transient charges/currents are investigated, revealing the strong influence on the device functionality originated from both the light-matter interactions in individual metasurface units and the dynamic coupling between them. Our work not only offers a flexible, reliable and cost-effective solution for chiral terahertz wave generation and manipulation, but also opens a new pathway to metasurface-tailored spintronic devices for efficient vector-control of electromagnetic waves in the terahertz regime.

Coherent terahertz sources driven by femtosecond laser pulses can now routinely generate sub-picosecond few-cycle terahertz waves with exceptionally stable carrier waveforms, which can be used in numerous fundamental studies and practical applications[1,2]. The ability to manipulate the three-dimensional (3D) electric-field vector of such broadband terahertz waveforms can substantially broaden the applications of the terahertz technologies, and open up new possibilities for studies of coherent light-matter interactions[3–6], as well as of novel ultrafast quantum control carrier dynamics facilitated by phase-stable strong terahertz fields[7–9]. Therefore, a great amount of research has been devoted to realize full control of terahertz electric-field vectors in their amplitude, phase, frequency, polarization and spatial properties. Hitherto, the developed methods can be categorized into 1) the direct generation from solid or gas-plasma sources, for example, by applying external fields[10–12], by pulse-shaping of the driving laser pulses[13] or by a combined two-color laser scheme[6,14,15] and 2) the implementation of passive optical components, such as terahertz polarizers[16], waveplates[17,18], as well as terahertz metasurfaces[19,20]. However, a flexible and robust solution for efficient vector-control of broadband terahertz waves is still elusive.

Notably, in the terahertz regime, metasurfaces have greatly enriched the capabilities and freedom of wave manipulation[21–23]. By engineering with light-matter interactions in two-dimensional (2D) subwavelength building blocks ("meta-atoms") and collective coupling between them, metasurfaces can precisely control the field transformation of incident wavefronts and achieve predesigned functionalities. More recently, utilizing nonlinear optical effects, direct generation of terahertz radiations from laser-excited metasurfaces was also demonstrated[24,25], opening the door for the integration of broadband terahertz emitters with metasurface technologies. This could lead to compact and flexible terahertz sources with the full access to waveform control. However, the generation efficiency is still quite low, due to the low nonlinearity of the metasurface units. Therefore, new materials and systems are under exploration to increase the nonlinear conversion efficiency.

This dilemma can be overcome by the spintronic terahertz emitters composed of ferromagnetic (FM) and non-magnetic (NM) heterostructures[26–28]. The spintronic emitters exhibit the advantages of being low-cost, highly reliable, efficient and flexible, allowing implementations with a wide range of driving laser conditions, from nJ pulse energy of a compact laser oscillator to mJ pulses from a laser amplifier. An ultrabroad spectral bandwidth covering 1-30 THz can be produced when excited by short 10-fs laser pulses[27]. In particular, because the micro-nano fabrication of metal thin-film heterostructures is technologically well established, these spintronic emitters can be easily made into various metasurface structures, opening up great potential for applications. When excited by femtosecond laser pulses, the laser-induced transient currents, which are inherent to the spintronic emitters, can serve as efficient and active driving sources of these metasurfaces, the properties of which can be well controlled by the excitation lasers and external fields. Hence, such a hybrid terahertz emitter has the potential for high-efficiency terahertz-wave generation and manipulation in a single device. Understanding the influence of the metasurface structure on the laser-induced charge and current dynamics on the microscopic level is the key for sophisticated device design in the future.

In this letter, we propose a novel spintronic-metasurface terahertz emitter, consisting of metasurface-patterned FM/NM heterostructures. Taking the prototypical stripe-pattern metasurface as an example, we demonstrate the generation and manipulation of chiral terahertz waveforms in an efficient and highly flexible manner. By simply varying the transient spintronic-origin currents with an orientated external magnetic field, the emitter functionality can be actively controlled, leading to continuous tuning of the terahertz polarization state and helicity. The interplay between the laser-induced driving currents and the metasurface-induced charges and currents are investigated experimentally and theoretically under different metasurface geometries, revealing the strong influence on the device functionality of both the light-matter interactions in individual metasurface units and the dynamical coupling over the entire metasurface. Our work opens a new pathway to metasurface-tailored spintronic devices for efficient control over the electric-field vectors in the terahertz regime.

**Results**

Figure 1a illustrates the schematic of the experimental setup. In our experiments, the ultrashort laser pulses (duration 26 fs, center wavelength 1030 nm and repetition rate 100 kHz) generated by high-efficiency pulse compression of a Yb:KGW laser amplifier are used to excite the active spintronic-metasurface device. The pulse compression is enabled by the solitary beam propagation in periodic layered Kerr media (PLKM)[29]. The excitation pulse energy is ~20 μJ and the beam radius on the metasurface emitter is ~1.1 mm. (see Methods and Supplementary Section S1 for the details of the experimental setup) The metasurface emitter is composed of stripe-patterned FM/NM heterostructures. The FM/NM heterostructure consists of NM Pt (thickness of 3nm), capped with FM $Fe_{50}Co_{50}$ (1.4 nm) and supported by a thick $SiO_2$ or $Al_2O_3$ substrate. The stripe patterns are then fabricated by the standard optical lithography and ion beam etching process (see Methods). The metasurface structures with different stripe widths ($d$) and spacings (*l*), as well as on different substrates (SiO$_2$, Al$_2$O$_3$), are investigated (Fig. 1b). In the experiments, the stripe orientation is fixed along the *x*-axis, while the magnetization of FM layer (**M**) is saturated by an oriented external magnetic field (**H**) with a field magnitude of 200 mT (see Supplementary Section S3), and its direction can be continuously adjusted in the *x*-*y* plane. The field angle $\theta_H$ is defined as the angle between **H** and the stripe orientation (*x*-axis) (Fig. 1a). The applied field is much stronger than the anisotropy field of the Fe film, thus the Fe magnetization is expected to always align parallel to **H**. Under femtosecond laser illumination, the longitudinal spin current (**j**$_s$) arising in FM layer is converted into a transverse charge current (**j**$_c$) via inverse spin-hall effect (ISHE) in the NM layer, given by $\mathbf{j}_c = \gamma \mathbf{j}_s \times \frac{\mathbf{M}}{\|\mathbf{M}\|}$ (Fig. 1b and c), where $\gamma$ is the spin-Hall angle of NM[28] and and $\|\ \|$ denotes the magnitude of a vector. As a result, in our experiments, the laser-induced current **j**$_c$ flows perpendicular to **H** and serves as an active driving source of the stripe-patterned metasurface. The emitted terahertz field and its polarization state are then detected by the polarization- and time-resolved terahertz spectroscopy setup based on electro-optic sampling (EOS)[30–32] (see Methods and Supplementary Section S2).

We first present evidence showing that the metasurface can influence the device functionality by inducing strong amplitude and phase modulations onto the emitted terahertz waveforms. The EOS signals for the terahertz-wave components polarized parallel ($E^{\parallel}$) and perpendicular ($E^{\perp}$) to the stripes are plotted in Fig. 2a. The results were obtained from a device with *d*=5 μm and the filling factor (FF) *d*/(*d*+*l*)=0.5 on a SiO$_2$ substrate, and similar observations can be made on other metasurface geometries (see Supplementary Sections S5-7). Clearly, the perpendicular electric-field amplitude $E^{\perp}$ is strongly suppressed compared to $E^{\parallel}$, which is consistent with the previous work[26,33]. The results of the peak-to-peak amplitude $V_{pp}$ (see Fig. 2a and b) as a function of $\theta_H$ for $E^{\parallel}$ and $E^{\perp}$ are summarized in Fig. 2c, both of which exhibit a sinusoidal behavior while a ~90° angle shift can be observed. Furthermore, our results in Fig. 2a and b show that the terahertz waveforms, respectively for the two orthogonal polarizations, possess almost identical temporal waveforms. This conclusion is further corroborated by the normalized spectra shown in Fig. 2d, which displays identical spectral shapes for each polarization, although the spectra of $E^{\perp}$ is blue-shifted compared to $E^{\parallel}$ [26,33]. The coherent detection of EOS allows us to retrieve the phase information and, most interestingly, the phase difference ($\varphi^{\perp} - \varphi^{\parallel}$) stays close to $\pm\pi/2$ throughout the entire spectrum (Fig. 2e). This clearly indicates the generation of chiral terahertz waves.

Our findings here are in distinct to past work, which only focused on the amplitude modulation and spectral blueshifts of the terahertz waves from the stripe-patterned terahertz emitters[26,33]. Instead, our results clearly show that the directions parallel and perpendicular to the stripes define a set of canonical coordinates, in which the terahertz waveforms of $E^{\parallel}$ and $E^{\perp}$ are decoupled from each other and possess a broadband quarter-wave phase difference. This is the key for the generation and manipulation of the chiral terahertz waves.

The above observations can be conceptually captured by a model, which considers the transients of the spintronics-origin current density ($\mathbf{j}_c$) and the metasurface-induced transient charges ($Q_i$) and current density ($j_i$) flowing in $y$ direction (see Fig. 1c). According to Ohm's law, the electric-field component $E^{\parallel}$ ($E^{\perp}$) of the emitted terahertz wave is proportional to the total current density flowing in the same direction, $j_a^{\parallel}$ ($j_a^{\perp}$), by $E^{\parallel,\perp}(\omega) = j_a^{\parallel,\perp}(\omega)/\sigma(\omega)$ [27,28], where $\sigma$ is the conductance of the metal layer. As shown in Fig. 1c, along the stripes ($x$ axis), the current density $j_a^{\parallel}$ is solely contributed by the *x*-component of $\mathbf{j}_c$, given by $j_a^{\parallel} = \|\mathbf{j}_c\| \sin\theta_H$. Here, $\mathbf{j}_c$ flows in the direction perpendicular to **H** (Fig. 1c)[28]. Perpendicular to the stripes (*y* axis), on the other hand, the current density $j_a^{\perp}$ consists of both the *y*-component of $\mathbf{j}_c$ and the metasurface-induced "counteractive" current $j_i$ (see Fig. 1c), which yields $j_a^{\perp} = -\|\mathbf{j}_c\|\cos\theta_H + j_i$. The counteractive current $j_i$ is driven by the electric field built up by the transient charge density $Q_i$ with $j_i = \sigma Q_i / C$, where *C* is the proportionality constant between the charge-induced electric field in the metal layer and the corresponding charge density, which is determined by the geometry of the metasurface structure (see Supplementary Section S4). Here, $Q_i$ can be considered as the result of the accretion of $j_a^{\perp}$ at the stripe boundaries. In the frequency domain, we derive that $j_i(\omega) = -\dfrac{\sigma(\omega)}{i\omega C} j_a^{\perp}(\omega)$. For convenience of discussion, we further assume the low-frequency limit ($\omega \to 0$) and the conductance $\sigma$ becomes a constant[34], which finally yields $E^{\parallel}(\omega) = \dfrac{\|\mathbf{j}_c(\omega)\|}{\sigma}\sin\theta_H$ and $E^{\perp}(\omega) = -\dfrac{i\omega C}{\sigma^2}\|\mathbf{j}_c(\omega)\|\cos\theta_H$ (see Supplementary Section S4 for detailed derivation).

First of all, we obtain from the above model that the amplitude of $E^{\perp}$ is scaled by a factor of $\omega C/\sigma$ ($\omega \to 0$) when compared to $E^{\parallel}$, which is consistent with the observed suppression of $E^{\perp}$ (Fig. 2a and b). This also explains the spectral blueshift of $E^{\perp}$ with respect to $E^{\parallel}$ (Fig. 2d). Second, the amplitudes of $E^{\parallel}$ and $E^{\perp}$ follow the sine- and cosine-functions of $\theta_H$, respectively, which is consistent with the results in Fig. 2c. Finally, the complex amplitudes of $E^{\parallel}$ and $E^{\perp}$ exhibit a spectral quarter-wave phase difference, which also agrees with the experimental results in Fig. 2e. We note that, although the simple model above can qualitatively explain the general features of our observations, the quantitative agreement over the entire spectrum is elusive. Furthermore, the inductive coupling between the transient currents is either not considered in this model. Hence, numerical simulations using the frequency-domain solver of COMSOL Multiphysics[35] are further conducted to provide a comprehensive understanding of our results and to extract the microscopic mechanisms (see Methods).

The broadband quarter-wave phase difference naturally leads to chiral terahertz emission. In Fig. 3a, we plot the experimentally measured time dependence of the electric-field vector for a chiral terahertz waveform, with the three projections displaying the waveforms of the mutually orthogonal components $E^{\parallel}(t)$ and $E^{\perp}(t)$, and their parametric plot. Here, a stripe pattern with $d$=25 μm and FF=0.5 is excited by the laser pulses under a field angle $\theta_H$ of 33º, which yields equal peak amplitudes of $E^{\parallel}$ and $E^{\perp}$ (See Supplementary Section S5). As illustrated in Fig. 3b, the ellipticity and handedness of the terahertz radiation can be conveniently and continuously tuned by changing the field angle $\theta_H$. Such tunability relies on the fact that we can independently adjust the intensities of the two orthogonal components, while the relative quart-wave phase difference can be maintained.

In Fig. 3c, we summarize the broadband ellipticity $\langle\varepsilon\rangle$ under different stripe widths $d$ for FF=0.5, accompanied by the relative intensity ($\eta$) of the terahertz fields compared to that from a homogeneous thin film with the same FM/NM heterostructure (see Methods). Here, the broadband ellipticity $\langle\varepsilon\rangle$ is measured under an optimum $\theta_H$ (see Supplementary Section S5) and characterized by considering the spectral intensity and the phase difference over the entire spectrum (see Methods). These results can be well reproduced by our numerical simulations (dashed lines in Fig. 3c). The experimental spectrum-resolved ellipticity $\varepsilon(\omega)$ is plotted in Fig. 3d. We find that chiral terahertz waves can be produced from metasurfaces with $d$<80 μm, while the circularity monotonically declines for wider stripes. This can be attributed to two factors: First, as $d$ increases, both of the metasurface-induced charge density ($Q_i$) and the counteractive current density $\mathbf{j}_i$ diminish, leading to the rapid deviation of $\left|\varphi^\perp - \varphi^\|\right|$ from $\pi/2$ at high spectral frequencies (see Supplementary Section S5). Second, as shown in Fig. 3d, plummets of the ellipticity can be observed beyond specific resonant frequencies (open symbols, and see below). These anomaly features result in a stronger influence to the field circularity when the stripes become wider, as their frequencies gradually shift and meet with the spectral peak of the terahertz waves (~1.5 THz). In our experiments, the broadband ellipticity $\langle\varepsilon\rangle$ can be as high as 0.75 in narrow stripes ($d$=3-10 μm), and this value is mostly restricted by the blueshift of the $\mathbf{E}^\perp$ spectrum (Fig. 2d). Almost perfect circular polarization ($\varepsilon\approx1$), on the other hand, can be realized in a narrower bandwidth between 1.5 to 2.5 THz. Owing to the strong transverse confinement of the narrow stripes, however, the corresponding terahertz intensity is generally one order of magnitude lower compared to the emission from a homogeneous thin film (Fig. 3c). Otherwise, stronger elliptically polarized terahertz waves with $\langle\varepsilon\rangle$ ~0.6 can be generated with a relative intensity of $\eta$~40 % at $d\approx20$ μm.

The resonant features in Fig. 3d can be attributed to the collective coupling dynamics between the transient charges and currents over the entire metasurface. In Fig. 4a and b, we plot the experimental spectral amplitudes of $E^\perp$ and $E^\|$, respectively, normalized by that obtained from a homogeneous film $E_{homo}$ (see Supplementary Section S2). Multiple spectral anomalies can be observed, manifested as peaks and steps in the normalized spectra (Fig. 4a and b). Sharp drops away from $\pi/2$ of the relative phase $\left|\varphi^\perp - \varphi^\|\right|$ can also be resolved at these anomaly frequencies (see Supplementary Section S5), which are responsible for the

observed plummets of field ellipticity shown in Fig. 3d. These anomaly features can be mostly well captured by our numerical simulations (Fig. 4c and d), whereas the high-frequency dips present in the simulation for $E^{\parallel}$ spectra (Fig. 4d) are too weak to be observed in our experiments. In Fig. 4e, we summarize the anomaly frequencies for different periods (*d+l*), including the results obtained from FF≠0.5 (colored symbols) and those from the emitters with an Al$_2$O$_3$ substrate (half-filled symbols) (see Supplementary Section S7). We find that these low- and high-frequency anomalies can be well fitted by $f_a^{1,2} = \frac{1}{n}\frac{v_c}{d+l}$, where *n* is the refractive indices of the media and $v_c$ the speed of light in vacuum. For the high-frequency anomaly $f_a^2$ (open symbols in Fig. 4a and b), the refractive index of the air ($n_{\text{air}}$) yields an excellent fit, while the low-frequency anomaly $f_a^1$ (solid symbols) can be fitted by the terahertz refractive indices of the substrates ($n_{\text{SiO}_2}$ or $n_{\text{Al}_2\text{O}_3}$) (Fig. 4e), indicating its origin from the dynamical coupling through the substrate layer. Here, the different shapes of the anomaly features could be attributed to the Fano-like coupling between the narrow-band Rayleigh diffraction anomaly and the broad-band surface plasmon excitation, which is different for the TE and TM polarizations[36,37].

With the help of the numerical simulations, we try to better understand the generation mechanism for chiral terahertz waves in the sub-wavelength scale. Taking *d*=50 μm and FF=0.5 as an example, we plot in Fig. 4f and g the space- and frequency-distributions of the transient current density $j_a$ in a single metal stripe, which is normalized by the driving current density $\|\mathbf{j}_c\|$. For the currents along the stripe ($j_a^{\parallel}$, Fig. 4g), the skin effect can be clearly observed and the current density is largest near the boundaries of the stripe (*y*=±25 μm). On the contrary, the current density of $j_a^{\perp}$ (Fig. 4f) is almost completely suppressed at the boundaries due to the conductivity discontinuity, forming a standing-wave-like current distribution along *y* across the stripe. Indeed, the appearance of an additional node, which corresponds to a higher-order standing current wave, start to appear when the frequency is $>v_c/2d$ (~3 THz in this case). In the low-frequency region, $j_a^\perp$ is almost completely subdued throughout the entire stripe, which is consistent with the low-frequency suppression of the terahertz emission. This can be understood by the fact that more transient charges ($Q_i$) tend to be accumulated at the boundaries when the frequency is low, leading to a stronger counteractive current ($j_i$), which suppresses the flowing of $j_a^\perp$. This observation confirms that the confinement of laser-induced transient currents in the stripe-patterned metasurface is responsible for the observed spectral and phase modulations, as well as for the generation of chiral terahertz waveforms. The spectral anomaly in Fig. 4a-d can also be clearly resolved as the sharp peaks or steps of the current density at the corresponding frequencies (Fig. 4f and g).

**Discussion**

To increase the generation efficiency of the chiral terahertz waves, an intuitive way is to increase FF for a narrow stripe width (*d*) by reducing the stripe spacing (*l*). However, our results indicate that the ellipticity can be deteriorated due to the stronger dynamical coupling between the stripes and the appearance of the spectral anomalies. As shown in Fig. 5a, the broadband ellipticity $\langle\varepsilon\rangle$ for *d*=50 μm on a SiO$_2$ substrate decreases monotonically for FF>0.4. This can be explained by a stronger capacitive coupling between the stripes, which results in weaker counteractive currents ($j_i$) and less confinement of the transverse currents (see Supplementary Section S4). This is supported by the simulation results of the geometrical factor *C*, which monotonically increases when the stripes become denser (Fig. 5a). On the other hand, $\langle\varepsilon\rangle$ does not keep increasing for a lower FF as the stripes become more isolated. We find this is influenced by the appearance of the low-frequency anomaly ($f_a^1$) around the spectral peaks of the terahertz waves. Figure 5b shows the broadband ellipticity $\langle \varepsilon \rangle$ obtained from our simulation under different stripe widths and FFs, which generally agrees with our experimental data (Fig. 5a and c). Notably, the optimum $\langle \varepsilon \rangle$ exists in the region of FF=0.3-0.4 for a number of different stripe widths. This optimum region is determined by the complex interplay between the transient charge/current dynamics and the frequency anomalies. Given that our terahertz waves approximately centered at ~1.5THz and the low-frequency anomaly is, on the other hand, given by $f_a^1 = \frac{1}{n_{SiO_2}} \frac{v_c}{d+l}$. The agreement of the two yields $d+l \approx 100$ μm (white dashed line in Fig. 5b) which is in good correspondence with the boundary for the optimum region of $\langle \varepsilon \rangle$. As a result, we conclude that the broadband ellipticity $\langle \varepsilon \rangle$ can be further increased by choosing a substrate with low terahertz refractive index and pushing the anomaly frequency beyond the spectral range of the terahertz wave.

The spintronics-metasurface emitter in our work represents a stable, highly tunable and economical solution for generating broadband elliptically polarized terahertz radiations. Considering that a peak field up to 300 kV cm$^{-1}$ can be generated from a homogeneous thin-film emitter when excited by a multi-millijoule laser amplifier[38], our approach has the potential to generate >100 kV cm$^{-1}$ chiral terahertz field under similar conditions with a broadband ellipticity $\langle \varepsilon \rangle$ ~0.6 (Fig. 3c). This field strength is already comparable to the chiral terahertz emission from laser-plasma sources[11]. Moreover, the same emitting device can also be compatible with a nJ-level laser oscillator[27], which highlights the great flexibility of our method. Finally, although the magnetic properties and the spintronic dynamics in the individual metasurface units are identical in this work, the capability of modern spintronic nanoscale engineering allows the manipulation of magnetization, magnetic anisotropy, spin-current dynamics in individual blocks. This offers a new degree of freedom to tailor the functionality of spintronic-metasurface devices, which could potentially lead to arbitrary vector-control of broadband terahertz waves in both space and time.

**Methods**

**Generation of ultrashort femtosecond pulses.** Short femtosecond pulses (duration 26 fs, center wavelength 1030nm) are generated by supercontinuum generation and pulse compression of the 170 fs pulses from a Yb:KGW amplifier laser system. The pulses with a repetition rate of 100 kHz and an initial energy of 80 μJ from the Yb:KGW amplifier are focused into the PLKM composed of 12 $SiO_2$ plates with the thickness of 800 μm and the spacing of 2.54 cm. The plates are placed at the Brewster angle to minimize the transmission loss. The output pulses are spectrally broadened and the dispersion is further compensated by a set of chirped mirrors. Ultrashort pulses with a duration of ~24 fs and an energy of 60 μJ are generated. To avoid sample damage, the energy is further attenuated to ~ 20 μJ for the excitation of the spintronic-metasurface emitters. About 10% of the energy is used as the ultrashort probe pulse for EOS. See Supplementary Section S1 for the details of the experimental setup.

**Sample fabrication.** The $Co_{50}Fe_{50}$/Pt heterostructures are grown on $SiO_2$ wafers with 2-inch diameter by confocal dc magnetron sputtering. The substrates are cleaned by a short plasma etching and the following annealing at 200°C for 10 minutes. A $Co_{50}Fe_{50}$ alloy target is used for growing the CoFe film. After the growth of the CoFe film, the sample is capped by a 5 nm

SiO$_2$ layer using rf sputtering. All the layers are grown at room temperature with the Ar pressure of 3 mTorr. The deposition rate is 0.55 Å/s for Pt layer and 0.30 Å/s for CoFe alloy layer. The thickness is 3 nm for the Pt layer and 1.4 nm for the CoFe layer, which are optimized for maxim THz radiation.

After the film growth, the sample is cut into small square pieces with the size of ~10 mm, and each piece of the SiO$_2$/CoFe/Pt sample is patterned into the microstrips with different period and filling factor using the standard optical lithography and ion beam etching process.

**Polarization-resolved terahertz time-domain spectroscopy.** The elliptically polarized terahertz waves can be decomposed into mutually orthogonal components $E^{\parallel}$ and $E^{\perp}$ (Fig. 1a). By measuring the terahertz electric fields polarized in two directions, we are able to recover the full three-dimensional profiles of the elliptically polarized terahertz waves and characterize their ellipticity. In our experiments, the terahertz waveforms are measured with the EOS method, and the fields polarized along the directions perpendicular and parallel to the stripes are resolved by implementing a terahertz wire-grid polarizer and adjusting the EOS crystal angle accordingly. For the detection of the terahertz amplitude, we use a 300-μm-thick GaP(110) crystal in the EOS setup. The polarization-resolved terahertz time-domain spectroscopy setup is characterized by the emission of a thin-film spintronic emitter under the oriented magnetic field, which yields linear polarization perpendicular to the magnetic field (see SM).

The ellipticity of the terahertz waves is characterized by

$$\langle \varepsilon \rangle = \sqrt{\frac{\int_0^{\infty} \varepsilon^2(\omega)\left[I^{\parallel}(\omega)+I^{\perp}(\omega)\right]d\omega}{\int_0^{\infty}\left[I^{\parallel}(\omega)+I^{\perp}(\omega)\right]d\omega}} \quad (1)$$

, where $I^{\parallel,\perp}(\omega) = \dfrac{v_c \varepsilon_0 \left[ E^{\parallel,\perp}(\omega) \right]^2}{2}$ is the spectral intensity, and $\varepsilon^2$ is given by[39]

$$\varepsilon^2 = \frac{\left(E^{\parallel}\right)^2 + \left(E^{\perp}\right)^2 - \sqrt{\left(\left(E^{\parallel}\right)^2 - \left(E^{\perp}\right)^2\right)^2 + 4\left(E^{\parallel}E^{\perp}\right)^2 \cos^2\left(\varphi^{\parallel} - \varphi^{\perp}\right)}}{\left(E^{\parallel}\right)^2 + \left(E^{\perp}\right)^2 + \sqrt{\left(\left(E^{\parallel}\right)^2 - \left(E^{\perp}\right)^2\right)^2 + 4\left(E^{\parallel}E^{\perp}\right)^2 \cos^2\left(\varphi^{\parallel} - \varphi^{\perp}\right)}} \quad (2)$$

. Here, $\langle \varepsilon \rangle = 1$ corresponds to full circular polarization. The relative intensity $\eta$ is given by

$$\eta = \frac{\int_0^{\infty} \left[I^{\parallel}(\omega) + I^{\perp}(\omega)\right] d\omega}{\int_0^{\infty} I_{\text{homo}}(\omega) d\omega} \quad (3)$$

, where $I_{\text{homo}}$ is the spectral intensity measured under the same conditions from a homogeneous thin-film emitter with an identical FM/NM heterostructure.

**Numerical simulation.** The numerical simulation in this work are obtained with the frequency-domain solver of COMSOL Multiphysics[39], a finite-element-method-based software. In our simulations, we use a non-dispersive conductivity for the modeling of a 5nm-thick Fe-Co layer, $\sigma_{\text{Fe-Co}} = 2 \times 10^6$ S/m, which is determined experimentally (see Supplementary Section S3). We impose the periodic boundary condition along the $x$ and $y$ directions and a perfectly-matched-layer (PML) on the $z$ direction for the outgoing-wave (non-reflective) boundary condition. The thickness of the substrate is set to be infinite, by which we ignore the subsequent terahertz signals caused by the multiple reflection of the electromagnetic waves at the upper and lower surfaces of the 200-um-thick substrate. The terahertz refractive indices of the substrates are $n_{\text{SiO}_2} \approx 1.95$ and $n_{\text{Al}_2\text{O}_3} \approx 3.07$ [40], respectively. The spin-to-charge conversion is modeled by imposing a transverse external current density $\mathbf{j}_c(\omega)$ uniformly distributed in the FM/NM heterostructure layer and the direction of the driving current is perpendicular to the magnetic field. The amplitude and phase of $\mathbf{j}_c(\omega)$ can be obtained from the experimental results of the terahertz emission from the homogeneous thin films (see Supplementary Section S2). The simulation results are the output of the amplitude and phase of the far field in air above the surface for various metasurface parameters. The simulation results in frequency domain are obtained by using a Fast-Fourier Transform method, and are compared with the experimental data.

**References**

1.  Ferguson, B. & Zhang, X. Materials for terahertz science and technology. *Nat. Mater.* **1**, 26–33 (2002).

2.  Tonouchi, M. Cutting-edge terahertz technology. *Nat. Photonics* **1**, 97–105 (2007).

3.  Ulbricht, R., Hendry, E., Shan, J., Heinz, T. F. & Bonn, M. Carrier dynamics in semiconductors studied with time-resolved terahertz spectroscopy. *Rev. Mod. Phys.* **83**, 543–586 (2011).

4.  Kampfrath, T., Tanaka, K. & Nelson, K. A. Resonant and nonresonant control over matter and light by intense terahertz transients. *Nat. Photonics* **7**, 680–690 (2013).

5.  Liu, M. *et al.* Terahertz-field-induced insulator-to-metal transition in vanadium dioxide metamaterial. *Nature* **487**, 345–348 (2012).

6.  Schlauderer, S. *et al.* Temporal and spectral fingerprints of ultrafast all-coherent spin switching. *Nature* **569**, 383–387 (2019).

7.  Langer, F. *et al.* Lightwave-driven quasiparticle collisions on a subcycle timescale. *Nature* **533**, 225–229 (2016).

8.  Zaks, B., Liu, R. B. & Sherwin, M. S. Experimental observation of electron-hole recollisions. *Nature* **483**, 580–583 (2012).

9. Reimann, J. *et al.* Subcycle observation of lightwave-driven Dirac currents in a topological surface band. *Nature* **562**, 396–400 (2018).

10. Su, Q., Xu, Q., Zhang, N., Zhang, Y. & Liu, W. Control of terahertz pulse polarization by two crossing DC fields during femtosecond laser filamentation in air. *J. Opt. Soc. Am. B* **36**, G1–G5 (2019).

11. Wang, W. M., Gibbon, P., Sheng, Z. M. & Li, Y. T. Tunable circularly polarized terahertz radiation from magnetized gas plasma. *Phys. Rev. Lett.* **114**, 253901 (2015).

12. Lu, X. & Zhang, X. C. Generation of elliptically polarized terahertz waves from laser-induced plasma with double helix electrodes. *Phys. Rev. Lett.* **108**, 123903 (2012).

13. Sato, M. *et al.* Terahertz polarization pulse shaping with arbitrary field control. *Nat. Photonics* **7**, 724–731 (2013).

14. Kanda, N. *et al.* The vectorial control of magnetization by light. *Nat. Commun.* **2**, 362 (2011).

15. Zhang, Z. *et al.* Manipulation of polarizations for broadband terahertz waves emitted from laser plasma filaments. *Nat. Photonics* **12**, 554–559 (2018).

16. Ferraro, A. *et al.* Flexible terahertz wire grid polarizer with high extinction ratio and low loss. *Opt. Lett.* **41**, 2009–2012 (2016).

17. Shan, J., Dadap, J. I. & Heinz, T. F. Circularly polarized light in the single-cycle limit: The nature of highly polychromatic radiation of defined polarization. *Opt. Express* **17**, 7431–7439 (2009).

18. Masson, J. & Gallot, G. Terahertz achromatic quarter-wave plate. *Opt. Lett.* **31**, 265–267 (2006).

19. Jia, M. *et al.* Efficient manipulations of circularly polarized terahertz waves with transmissive metasurfaces. *Light Sci. Appl.* **8**, 16 (2019).

20. Grady, N. K. *et al.* Terahertz metamaterials for linear polarization conversion and anomalous refraction. *Science* **340**, 1304–1307 (2013).

21. He, Q., Sun, S., Xiao, S. & Zhou, L. High-Efficiency Metasurfaces: Principles, Realizations, and Applications. *Adv. Opt. Mater.* **6**, 1800415 (2018).

22. Pendry, J. B., Schurig, D. & Smith, D. R. Controlling electromagnetic fields. *Science* **312**, 1780–1782 (2006).

23. Hsiao, H.-H., Chu, C. H. & Tsai, D. P. Fundamentals and Applications of Metasurfaces. *Small Methods* **1**, 1600064 (2017).

24. Li, G., Zhang, S. & Zentgraf, T. Nonlinear photonic metasurfaces. *Nat. Rev. Mater.* **2**, 1–14 (2017).

25. McDonnell, C., Deng, J., Sideris, S., Ellenbogen, T. & Li, G. Functional THz emitters based on Pancharatnam-Berry phase nonlinear metasurfaces. *Nat. Commun.* **12**, 30 (2021).

26. Yang, D. *et al.* Powerful and Tunable THz Emitters Based on the Fe/Pt Magnetic Heterostructure. *Adv. Opt. Mater.* **4**, 1944–1949 (2016).

27. Seifert, T. *et al.* Efficient metallic spintronic emitters of ultrabroadband terahertz radiation. *Nat. Photonics* **10**, 483–488 (2016).

28. Kampfrath, T. *et al.* Terahertz spin current pulses controlled by magnetic heterostructures. *Nat. Nanotechnol.* **8**, 256–260 (2013).

29. Zhang, S. *et al.* Solitary beam propagation in periodic layered Kerr media enables

high-efficiency pulse compression and mode self-cleaning. *Light Sci. Appl.* **10**, 53 (2021).

30. Planken, P. C. M., Nienhuys, H.-K., Bakker, H. J. & Wenckebach, T. Measurement and calculation of the orientation dependence of terahertz pulse detection in ZnTe. *J. Opt. Soc. Am. B* **18**, 313 (2001).

31. Wu, Q. & Zhang, X. C. Free-space electro-optic sampling of terahertz beams. *Appl. Phys. Lett.* **67**, 3523 (1995).

32. Leitenstorfer, A., Hunsche, S., Shah, J., Nuss, M. C. & Knox, W. H. Detectors and sources for ultrabroadband electro-optic sampling: Experiment and theory. *Appl. Phys. Lett.* **74**, 1516–1518 (1999).

33. Jin, Z. *et al.* Terahertz Radiation Modulated by Confinement of Picosecond Current Based on Patterned Ferromagnetic Heterostructures. *Phys. Status Solidi - Rapid Res. Lett.* **13**, 1900057 (2019).

34. Nadvorník, L. *et al.* Broadband terahertz probes of anisotropic magnetoresistance disentangle extrinsic and intrinsic contributions. *Phys. Rev. X* **11**, 021031 (2020).

35. https://www.comsol.com/.

36. Liu, F. & Zhang, X. Fano coupling between Rayleigh anomaly and localized surface plasmon resonance for sensor applications. *Biosens. Bioelectron.* **68**, 719–725 (2015).

37. Savoia, S. *et al.* Surface sensitivity of Rayleigh anomalies in metallic nanogratings. *Opt. Express* **21**, 23531 (2013).

38. Seifert, T. *et al.* Ultrabroadband single-cycle terahertz pulses with peak fields of 300 kV cm-1 from a metallic spintronic emitter. *Appl. Phys. Lett.* **110**, 252402 (2017).

39. Fleischer, A., Kfir, O., Diskin, T., Sidorenko, P. & Cohen, O. Spin angular momentum and tunable polarization in high-harmonic generation. *Nat. Photonics* **8**, 543–549 (2014).

40. Palik, E. D. *Handbook of optical constants of solids*. **3**, (Academic press, 1998).

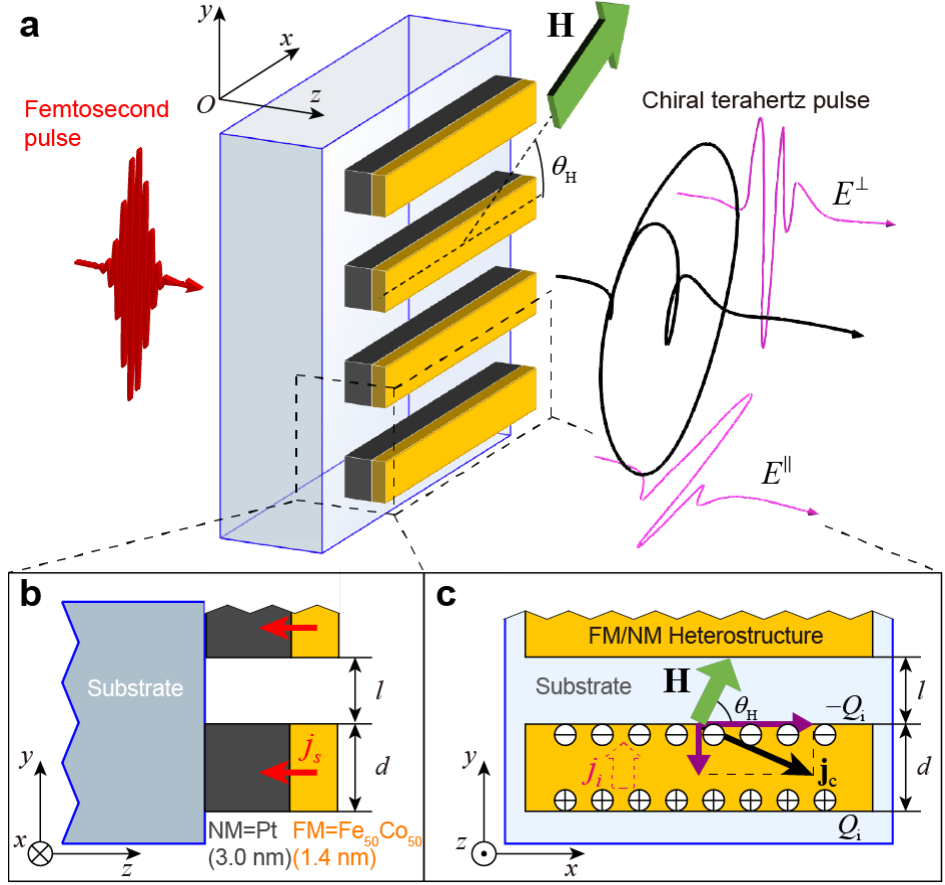

**Figure 1. Generation of chiral terahertz waves from a stripe-patterned spintronic-metasurface emitter. a.** Schematic of the experimental setup. An 1030nm laser pulse of 24 fs duration is focused to excite a stripe-patterned spintronic-metasurface emitter along the *z* direction. The stripe is aligned along the *x* direction. An orientated external magnetic field (**H**) is applied in the *x-y* plane with a field angle of *θ*. A few-cycle chiral terahertz pulse is generated, which can be decomposed into electric-field components polarized parallel ( $E^{\parallel}$ ) and perpendicular ( $E^{\perp}$ ) to the stripes. **b.** The illustration of the emitter structure in the *y-z* plane. The FM/NM heterostructure is composed of FeCo alloy of 1.4 nm and Pt of 3.0nm. The stripe width is *d* and the spacing between the stripes is *l*. Under laser illumination, spin currents **j**$_s$ are driven through the FM/NM interface. **c.** The illustration of the emitter structure in the *x-y* plane. Owing to ISHE, spin currents are converted to charge currents **j**$_c$ which flows perpendicular to **H**. The transverse confinement of stripes leads to charge accumulation (±$Q_i$) at the stripe boundaries.

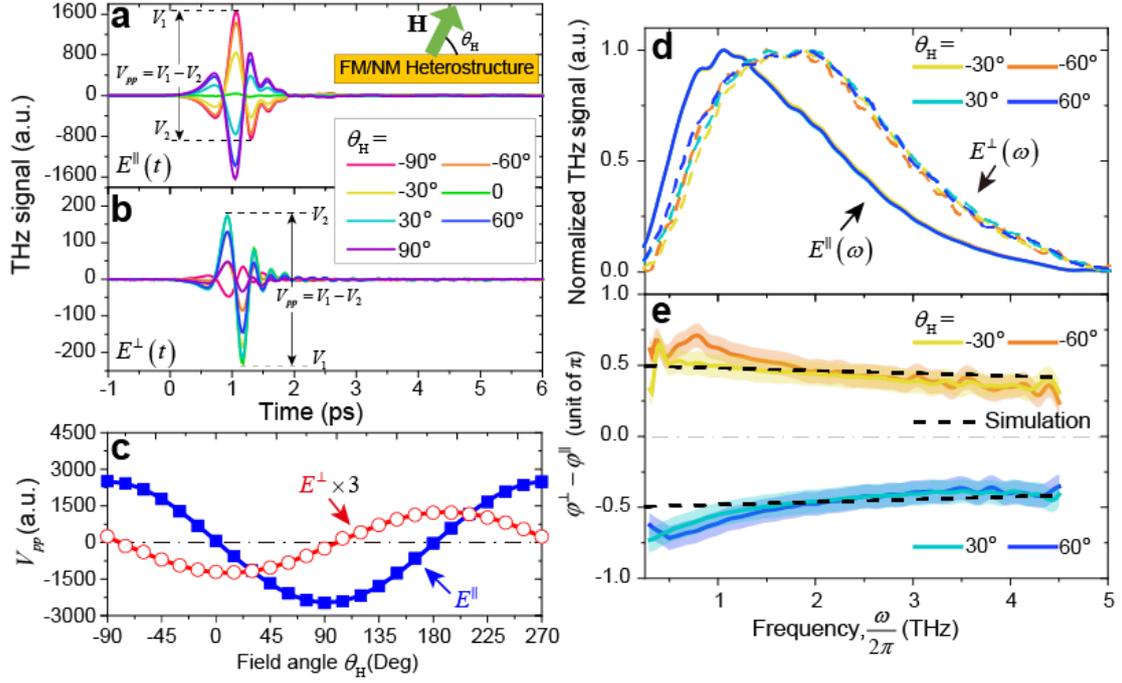

**Figure 2. Modulation of terahertz spectrum and phase due to metasurface structure. a.** Terahertz waveforms of $E^{\parallel}$ under different field angles $\theta_H$. The peak-to-peak field amplitude ($V_{pp}$) is defined as $V_1$-$V_2$. **b.** Same as **a** for $E^{\perp}$. **c.** The peak-to-peak field amplitude ($V_{pp}$) of $E^{\parallel}$ and $E^{\perp}$ (defined in **a** and **b**) as a function of $\theta_H$. **d.** Normalized spectra of $E^{\parallel}$ and $E^{\perp}$ under different field angles $\theta_H$. **e.** The relative phase difference between the parallel and perpendicular components $\varphi^{\perp} - \varphi^{\parallel}$ under different field angles $\theta_H$.

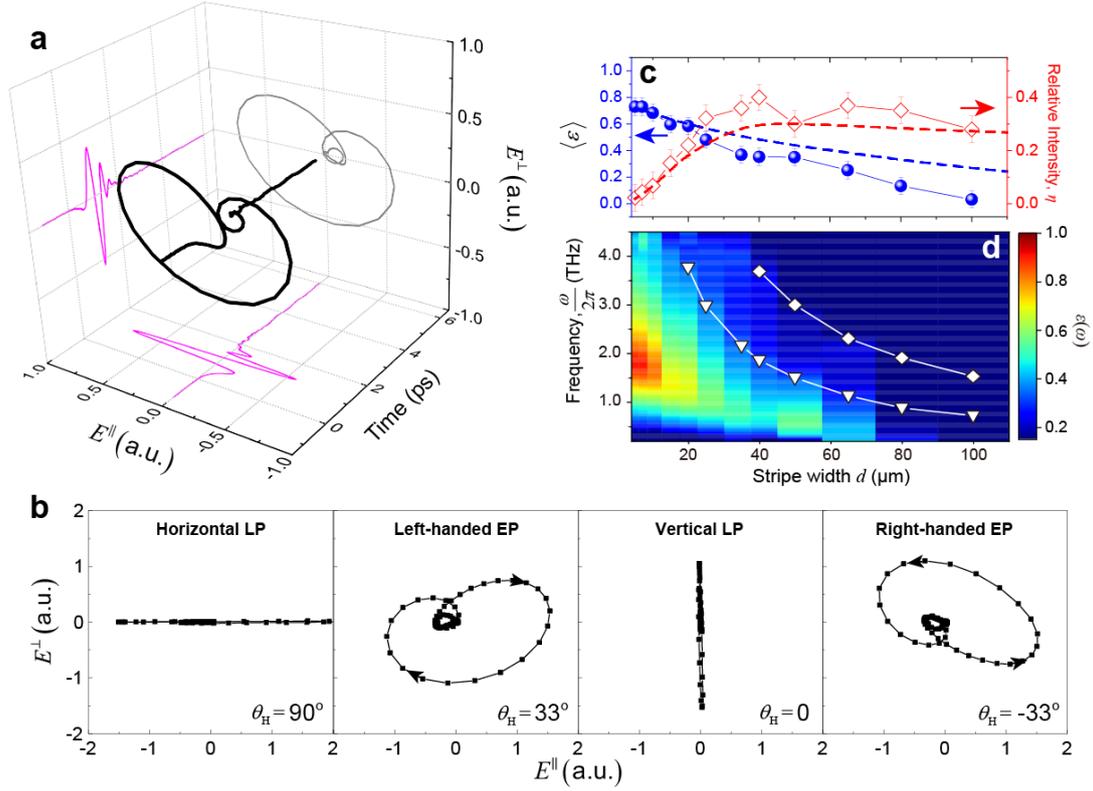

**Figure 3. Generation and manipulation of chiral terahertz waveforms. a.** Typical time dependence of the electric-field vector for a chiral terahertz waveform generated from a metasurface emitter with $d=l=25$ μm at a field angle $\theta_H=33°$. The three projections display the waveforms of the mutually orthogonal components $E^{\parallel}(t)$ and $E^{\perp}(t)$, and their parametric plot. **b.** The parametric plots of $E^{\parallel}(t)$ and $E^{\perp}(t)$ under different field angles $\theta_H$ under the same excitation conditions of **a.**, which illustrates the manipulation of terahertz polarization state and handedness with the orientation of **H**. LP: linear polarization; EP: elliptical polarization. **c.** The average ellipticity $\langle\varepsilon\rangle$ and the relative intensity $\eta$ for the chiral terahertz waves generated with different stripe width $d$. **d.** The spectrally resolved ellipticity $\varepsilon(\omega)$ as a function of $d$. The white filled symbols represent the anomaly frequencies ($f_a$) shown in Fig. 4e.

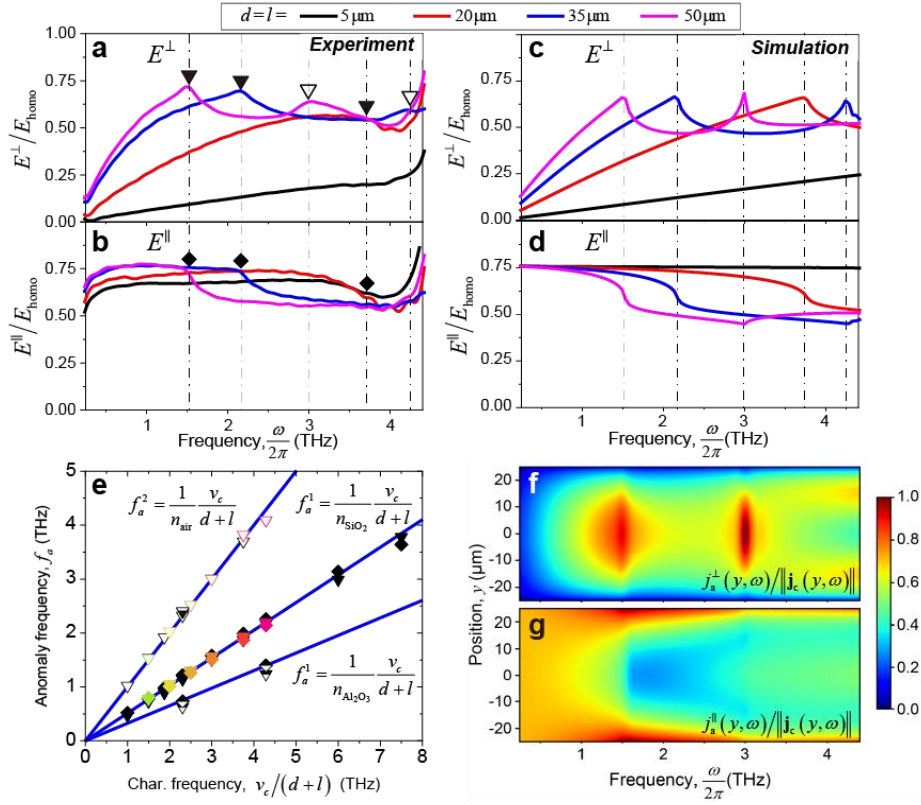

**Figure 4. Spectral anomaly due to coupling over the metasurface structure. a.** The normalized field spectra of $E^\perp$ for different stripe widths $d$ and spacings $l$ on a SiO$_2$ substrate measured in experiments. The spectra are normalized by those obtained from homogeneous thin-film emitters ($E_{\text{homo}}$). The solid triangles label the low-frequency anomaly features and the open triangles label the high-frequency ones. **b.** Same as a. for $E^\parallel$. The solid diamonds label the corresponding low-frequency anomaly features. **c.** and **d.** Simulation results obtained under the same conditions as **a** and **b**. The dash-dot lines align the corresponding anomaly features shared by $E^\parallel$ and $E^\perp$ spectra. **e.** The summary of the anomaly frequencies under different $d$ and $l$. The black filled triangles and diamonds represent the low-frequency features of $E^\perp$ and $E^\parallel$, respectively, for FF=50% (same as **a** and **b**). The black open triangles represent the high-frequency features of $E^\perp$ for FF=50%. The colored filled and open symbols are obtained from experiments with FF≠50% and the half-filled symbols illustrate anomaly frequencies measured from emitters on an Al$_2$O$_3$ substrate. **f.** The spatial and frequency distribution of the normalized total current density flowing perpendicular to the stripes $j_a^\perp$ for $d=l=50$ μm. **g.** same as **f** for $j_a^\parallel$.

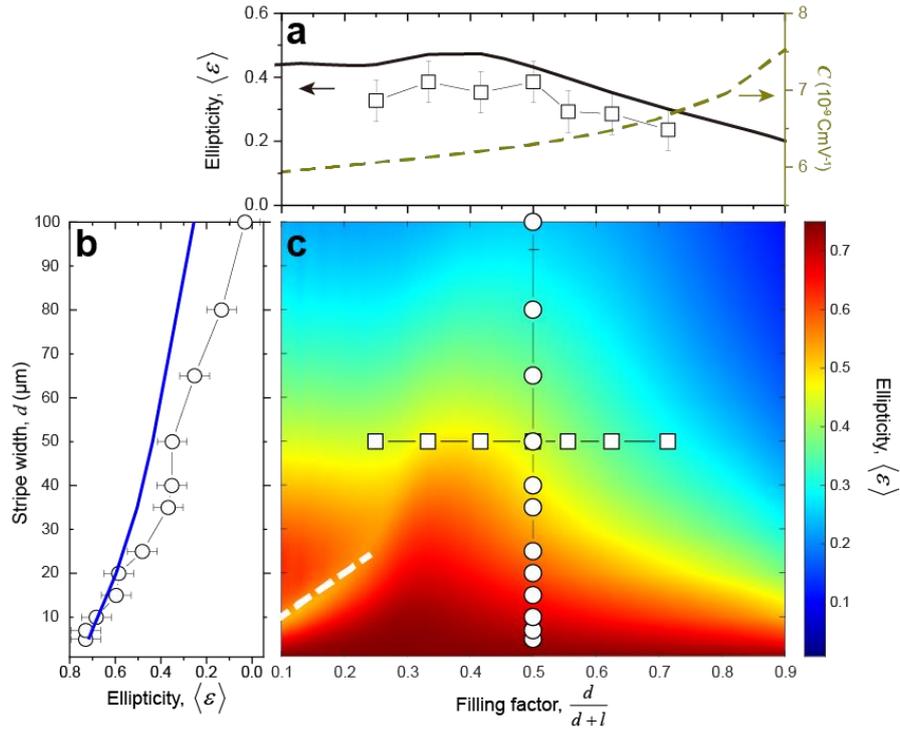

**Figure 5. Optimizing the ellipticity of spintronic-metasurface terahertz emitter. a**. The broadband ellipticity $\langle\varepsilon\rangle$ for $d$=50 μm as a function of the FF. The solid line is the simulation results. The simulation results of the geometrical factor $C$ as a function of FF are shown as the dashed line. **b.** $\langle\varepsilon\rangle$ for FF=50% as a function of the stripe width $d$. The solid blue line is the simulation results. **c.** 2D map of $\langle\varepsilon\rangle$ under different FFs and **d**. The open symbols label the experimental results in **a** and **b**. The dashed line corresponds to $d+l$=100 μm.